\documentclass[twocolumn,floatfix,showpacs,eqsecnum,prb]{revtex4}
\usepackage{graphicx}
\usepackage{amsmath}
\usepackage{bm}
\begin{document}

\title{Voltage probe model of spin decay in a chaotic quantum dot, with applications to spin-flip noise and entanglement production}
\author{B. Michaelis and C. W. J. Beenakker}
\affiliation{Instituut-Lorentz, Universiteit Leiden, P.O. Box 9506, 2300 RA Leiden, The Netherlands}
\date{December 2005}
\begin{abstract}
The voltage probe model is a model of incoherent scattering in quantum transport. Here we use this model to study the effect of spin-flip scattering on electrical conduction through a quantum dot with chaotic dynamics. The spin decay rate $\gamma$ is quantified by the correlation of spin-up and spin-down current fluctuations (spin-flip noise). The resulting decoherence reduces the ability of the quantum dot to produce spin-entangled electron-hole pairs. For $\gamma$ greater than a critical value $\gamma_{c}$, the entanglement production rate vanishes identically. The statistical distribution $P(\gamma_{c})$ of the critical decay rate in an ensemble of chaotic quantum dots is calculated using the methods of random-matrix theory. For small $\gamma_{c}$ this distribution is $\propto\gamma_{c}^{-1+\beta/2}$, depending on the presence ($\beta=1$) or absence ($\beta=2$) of time-reversal symmetry. To make contact with experimental observables, we derive a one-to-one relationship between the entanglement production rate and the spin-resolved shot noise, under the assumption that the density matrix is isotropic in the spin degrees of freedom. Unlike the Bell inequality, this relationship holds for both pure and mixed states. In the tunneling regime, the electron-hole pairs are entangled if and only if the correlator of parallel spin currents is at least twice larger than the correlator of antiparallel spin currents.
\end{abstract}
\maketitle

\section{Introduction}
\label{intro}

The voltage probe model, introduced by B\"{u}ttiker in the early days of mesoscopic physics,\cite{But88} gives a phenomenological description of the loss of phase coherence in quantum transport. Electrons that enter the voltage probe are reinjected into the conductor with a random phase, so they can no longer contribute to quantum interference effects. Such a device is no substitute for a microscopic treatment of specific mechanisms of decoherence, but it serves a purpose in identifying model independent ``universal'' features of the transition from coherent to incoherent electrical conduction. 

In this work we introduce and analyze a novel application of the voltage probe model, to spin-resolved conduction through a quantum dot. The voltage probe then serves a dual role: It randomizes the phase, as in the original spin-independent model,\cite{But88} but it also randomizes the spin. Two spin transport effects are examined: spin-flip noise and spin entanglement. The two effects are fundamentally connected, in the sense that the degree to which spin-up and spin-down current fluctuations are correlated provides a measure of the degree of spin entanglement of electron-hole pairs exiting the quantum dot.\cite{Bee04}

\begin{figure}
\centerline{\includegraphics[width=6cm]{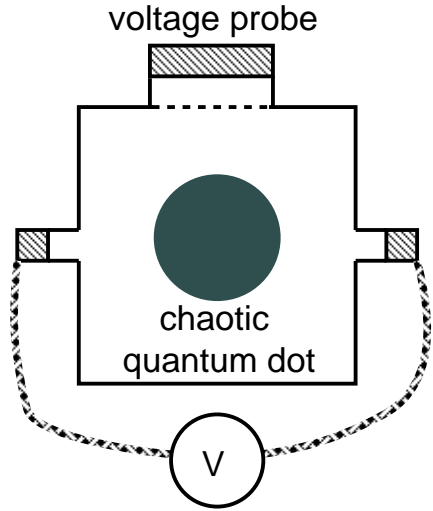}}
\caption{
Illustration of the voltage probe model. A chaotic quantum dot is connected to a voltage source by two single-channel leads. Decoherence is introduced by means of a fictitious voltage probe, which conserves particle number within each energy range $\delta E\ll eV$, on time scales $\delta t=h/\delta E$. (The dashed line in the figure indicates the tunnel barrier that separates the voltage probe from the quantum dot.) The random spin flips introduced by the voltage probe give a nonzero correlator of spin-up and spin-down currents (= spin-flip noise). The voltage probe also reduces the entanglement production by transforming the pure spin-singlet state of electron-hole pairs into a mixed Werner state.}
\label{fig_probe}
\end{figure}

The geometry is sketched in Fig.\ \ref{fig_probe}. The coupling of the electron spin to other degrees of freedom (nuclear spins, magnetic impurities, other electrons \ldots) is replaced by an artificial reservoir connected to the quantum dot via a tunnel barrier. The reservoir draws neither particles nor energy from the quantum dot.\cite{Pra05} Both the time averaged current and the time dependent fluctuations vanish, enforced by a fluctuating distribution function of the artificial reservoir.\cite{Bee92,DeJ96} This phenomenological description of decoherence has found many applications in the context of (spin-independent) shot noise. (Recent references include \onlinecite{Cle04,Chu05,Pol05,Pil05}.) For alternative models of decoherence in that context, see Refs.\ \onlinecite{Cle04,See01,Mar04,Mar05,Bee05}.

In the context of spin-resolved conduction, the voltage probe introduces two altogether different decay processes: spin flip and decoherence. These are characterized in general by two independent decay times (denoted $T_{1}$ and $T_{2}\leq 2T_{1}$, respectively). In order to obtain two different time scales we could introduce, in addition to the spin-isotropic voltage probe, a pair of ferromagnetic voltage probes that randomize the phase without flipping the spin (pure dephasing). Here we will restrict ourselves to the simplest model of a single voltage probe, corresponding to the limit $T_{2}=2T_{1}$. This choice is motivated by the desire to have as few free parameters as possible in this exploratory study. The more general model will be needed to make contact with the existing microscopic theory for the spin decay times.\cite{Gol04,Coi05,Bor05,Wit05}

The applications of the voltage probe model that we consider center around the concept of electron-hole entanglement. A voltage $V$ applied over a single-channel conductor produces spin-entangled electron-hole pairs.\cite{Bee03} The entanglement production rate is maximally $eV/2h$ bits/second, for phase-coherent spin-independent scattering. Thermal fluctuations in the electron reservoirs\cite{review} as well as dephasing voltage fluctuations in the electromagnetic environment\cite{Vel03,Sam03t} reduce the degree of entanglement of the electron-hole pairs. Unlike other quantum interference effects, which decay smoothly to zero, the entanglement production rate vanishes identically beyond a critical temperature or beyond a critical decoherence rate.

One goal of this investigation is to determine the probability distribution of the critical decoherence rate in an ensemble of quantum dots with chaotic scattering. The fluctuations in the artificial reservoir reduce the entanglement production by transforming the pure state of the electron-hole pair into a mixed state. For decoherence rates $\gamma\geq\gamma_{c}$ the density matrix of the electron-hole pairs becomes separable. The value of $\gamma_{c}$ is sample specific, with a probability distribution $P(\gamma_{c})$ that we calculate using the methods of random-matrix theory.\cite{Bee97}

The entanglement production is related to physical observables via the spin-resolved shot noise. The correlator of spin-up and spin-down currents (spin-flip noise) is of particular interest, since it provides a direct measure of the spin relaxation time.\cite{Sau04} By assuming that the elastic scattering in the quantum dot is spin-independent (no spin-orbit interaction), we derive a one-to-one relation between the degree of entanglement (concurrence) of the electron-hole pairs and the spin-resolved shot noise. In the more general spin-dependent case such a relation exists for pure states,\cite{Cht02,Sam04} through a Bell inequality, but not for mixed states.\cite{review} The expressions for the concurrence $\cal C$ take a particularly simple form in the tunneling regime, where we find that ${\cal C}$ is nonzero if and only if the correlator of spin current fluctuations is {\em at least twice larger for parallel spins than it is for antiparallel spins}.

We derive closed-form expressions for the ensemble averaged correlators in the regime of weak decoherence, both in the presence ($\beta=1$) and absence ($\beta=2$) of time-reversal symmetry. While the average spin-resolved current correlators are analytic in the decoherence rate $\gamma$ around $\gamma=0$, the average concurrence $\langle\cal C\rangle$ has a singularity at that point:  a square-root singularity $1-\langle {\cal C}\rangle\propto\sqrt{\gamma}$ for $\beta=1$, and a logarithmic singularity $1-\langle {\cal C}\rangle\propto\gamma|\ln\gamma|$ for $\beta=2$. The singular effect of a small decoherence rate on the entanglement production also shows up in the probability distribution $P(\gamma_{c})$ of the critical decoherence rate: It does not vanish for $\gamma_{c}\rightarrow 0$, but instead has a large weight $\propto\gamma_{c}^{-1+\beta/2}$.

The outline of this paper is as follows. We start in Sec.\ \ref{formulation} with a description of the system (quantum dot with voltage probe) and a formulation of the twofold question that we would like to answer (what is the entanglement production and how is it related to spin noise). A solution in general terms is presented in Sec.\ \ref{solution}. We begin in Sec.\ \ref{simplification} by simplifying the problem through the assumption of spin-independent scattering in the quantum dot. The concurrence of the electron-hole pairs is then given as a rational function of spin-resolved current correlators (Sec.\ \ref{currentcorr}). These correlators are expressed in terms of the scattering matrix elements of the quantum dot with voltage probe (Sec.\ \ref{smatrices}). To evaluate these expressions an alternative formulation, in terms of a quantum dot without voltage probe but with an imaginary potential, is more convenient (Sec.\ \ref{imaginary}). The connection to random-matrix theory is made in Sec.\ \ref{RMT}. By averaging over the random scattering matrices we obtain the nonanalytic $\gamma$-dependencies mentioned above (Secs.\ \ref{average} and \ref{critical}). We conclude in Sec.\ \ref{theend}.

\section{Formulation of the problem}
\label{formulation}

We consider a quantum dot coupled to source and drain by single-channel point contacts. The voltage probe has $N_{\phi}$ channels, and is connected to the quantum dot by a barrier with a channel-independent tunnel probability $\Gamma_{\phi}$. By taking the limit $\Gamma_{\phi}\rightarrow 0$, $N_{\phi}\rightarrow\infty$ at fixed (dimensionless) conductance $\gamma \equiv N_{\phi}\Gamma_{\phi}$, we model spatially homogeneous decoherence with coherence time \cite{Bro97}
\begin{equation}
\tau_{\rm coherence}=\lim_{\Gamma_{\phi}\rightarrow 0}\lim_{N_{\phi}\rightarrow\infty}\frac{h}{\gamma \Delta}.\label{tauphidef}
\end{equation}
(We denote by $\Delta$ the mean spacing of spin-degenerate levels.) Since the mean dwell time in the quantum dot (without voltage probe) is $\tau_{\rm dwell}=h/2\Delta$, one has 
\begin{equation}
\gamma=2\tau_{\rm dwell}/\tau_{\rm coherence}.\label{gammataudwell}
\end{equation}

The scattering matrix $S$ of the whole system has dimension $(N_{\phi}+2)\times(N_{\phi}+2)$. By convention the index $n=1$ labels the source, the index $n=2$ labels the drain, and the indices $3\leq n\leq N_{\phi}+2$ label the channels in the voltage probe. All of this refers to a single spin degree of freedom. Each channel is spin degenerate. As mentioned above, we assume that the scattering is spin independent. In particular, both the Zeeman energy and the spin-orbit coupling energy should be sufficiently small that spin rotation symmetry is not broken. The applied voltage $V$ between source and drain is assumed to be large compared to the temperature, but sufficiently small that the energy dependence of the scattering can be neglected.

The energy range $eV$ above the Fermi level is divided into small intervals $\delta E\ll eV$. The voltage probe conserves particle number and energy within each energy interval, on time scales $\delta t=h/\delta E$. We write this requirement as $I_{\phi}(E,t)=0$, where $I_{\phi}(E,t)$ is the electrical current through the voltage probe in the energy interval $(E,E+\delta E)$, averaged over the time interval $(t,t+\delta t)$.

Because the voltage probe does not couple different energy intervals, we may consider the entanglement production in each interval separately and sum over the intervals at the end of the calculation. In what follows we will refer to a single energy interval (without writing the energy  argument explicitly).

The density matrix $\rho$ of the outgoing state in each energy interval, traced over the degrees of freedom of the voltage probe, contains combinations of 0, 1, or 2 excitations in the spin degenerate channel of the source lead and the drain lead. Only the projection $\rho_{\rm eh}$ onto a singly excited channel in the source as well as in the drain contributes to the entanglement production.\cite{review} We denote by $w={\rm Tr}\,{\cal P}\rho$ the weight of the projection, with ${\cal P}$ the operator that projects onto singly excited channels (so that $w\rho_{\rm eh}={\cal P}\rho{\cal P}^{\dagger}$). The label ``eh'' stands for ``electron-hole pair'', where ``electron'' refers to the single excited channel in the drain and ``hole'' refers to the single nonexcited channel in the source.

In the absence of decoherence $\rho_{\rm eh}$ is pure ($\rho_{\rm eh}^{2}=\rho_{\rm eh}$). The voltage probe transforms $\rho_{\rm eh}$ into a mixed state. Our aim is to calculate the loss of entanglement of $\rho_{\rm eh}$ as a function of $\gamma$ and to relate it to the spin-resolved current correlators.

\section{General solution}
\label{solution}

\subsection{Simplification for spin-isotropic states}
\label{simplification}

The assumption of spin-independent scattering implies that the $4\times 4$ density matrix $\rho_{\rm eh}$ is invariant under the transformation $(U\otimes U)\rho_{\rm eh}(U^{\dagger}\otimes U^{\dagger})=\rho_{\rm eh}$, for any $2\times 2$ unitary matrix $U$. As a consequence, $\rho_{\rm eh}$ must be of the Werner form,\cite{Wer89}
\begin{equation}
\rho_{\rm eh}=\tfrac{1}{4}(1-\xi)\openone+\xi|\Psi_{\rm Bell}\rangle\langle\Psi_{\rm Bell}|,\;\; -\tfrac{1}{3}\leq \xi\leq 1,\label{rhoWerner}
\end{equation}
with $\openone$ the unit matrix and 
\begin{equation}
|\Psi_{\rm Bell}\rangle=2^{-1/2}\bigl(|\!\uparrow\downarrow\rangle-|\!\downarrow\uparrow\rangle\bigr)\label{Belldef}
\end{equation}
the Bell state.\cite{note3} (The spin-up and spin-down arrows $\uparrow,\downarrow$ label the two eigenstates of the Pauli matrix $\sigma_{z}$.)

The concurrence\cite{Woo98} (degree of entanglement) of the Werner state is given by
\begin{equation}
{\cal C}=\frac{3}{2}\max\left\{0,\xi-\frac{1}{3}\right\}.\label{CWerner}
\end{equation}
The Werner state is separable for $\xi\leq 1/3$. The entanglement production rate (bits per second) in the energy range $\delta E$ under consideration is given by\cite{review}
\begin{eqnarray}
{\cal E}&=&\frac{\delta E}{h}w{\cal F}\left(\tfrac{1}{2}+\tfrac{1}{2}\sqrt{1-{\cal C}^{2}}\right),\label{ECrelation}\\
{\cal F}(x)&=&-x\log_{2} x-(1-x)\log_{2}(1-x).\label{calFdef}
\end{eqnarray}

The parameter $\xi$ that defines the Werner state (\ref{rhoWerner}) can be obtained from the spin-spin correlator
\begin{equation}
{\rm Tr}\,\rho_{\rm eh}\sigma_{z}\otimes\sigma_{z}=-\xi.\label{Cxirelation}
\end{equation}
In order to make contact with the voltage probe model we now relate this correlator to a spin-resolved current correlator, along the lines of Refs.\ \onlinecite{Bee04,Cht02,Sam04}.

\subsection{Solution in terms of current correlators}
\label{currentcorr}

We define $N_{X,\alpha}^{\rm out}(t)$ as the number of electrons going out of the quantum dot in a time interval $(t,t+\delta t)$ through the source lead ($X=S$) or through the drain lead ($X=D$) with spin up ($\alpha=\uparrow$) or with spin down ($\alpha=\downarrow$). In terms of the current $I_{X,\alpha}(t)$ one has $N_{D,\alpha}^{\rm out}(t)=-I_{D,\alpha}(t)\delta t/e$, $N_{S,\alpha}^{\rm out}(t)=1-I_{S,\alpha}(t)\delta t/e$, with the convention that the current is positive if electrons enter the quantum dot. The spin-spin correlator (\ref{Cxirelation}) is expressed by
\begin{eqnarray}
\xi&=&-\frac{\langle [N_{S,\uparrow}^{\rm out}(t)-N_{S,\downarrow}^{\rm out}(t)][N_{D,\uparrow}^{\rm out}(t)-N_{D,\downarrow}^{\rm out}(t)]\rangle}{\langle [N_{S,\uparrow}^{\rm out}(t)+N_{S,\downarrow}^{\rm out}(t)][N_{D,\uparrow}^{\rm out}(t)+N_{D,\downarrow}^{\rm out}(t)]\rangle}\nonumber\\
&=&-(\delta t/e)^{2}\frac{1}{w}\langle [I_{S,\uparrow}(t)-I_{S,\downarrow}(t)][I_{D,\uparrow}(t)-I_{D,\downarrow}(t)]\rangle,\nonumber\\
&&\label{CNrelation}\\
w&=&(\delta t/e)^{2}\langle [I_{S,\uparrow}(t)+I_{S,\downarrow}(t)-2e/\delta t][I_{D,\uparrow}(t)+I_{D,\downarrow}(t)]\rangle,\nonumber\\
&&\label{wNrelation}
\end{eqnarray}
where the brackets $\langle\cdots\rangle$ indicate an average over many measurements.

The time dependent current $I_{X,\alpha}(t)=\bar{I}_{X,\alpha}+\delta I_{X,\alpha}(t)$ has time average $\bar{I}_{X,\alpha}$. The current fluctuations $\delta I_{X,\alpha}(t)$ on the time scale $\delta t=h/\delta E$ have cross correlator $\langle \delta I_{S,\alpha}(t)\delta I_{D,\beta}(t)\rangle=(\delta E/h)P_{\alpha\beta}$, with spectral density\cite{note1}
\begin{equation}
P_{\alpha\beta}=\int_{-\infty}^{\infty}dt\,\langle \delta I_{S,\alpha}(0)\delta I_{D,\beta}(t)\rangle.\label{Palphabetadef}
\end{equation}
The total spectral density of charge noise is given by
\begin{equation}
P_{\rm charge}=\sum_{\alpha,\beta}\int_{-\infty}^{\infty}dt\,\langle \delta I_{S,\alpha}(0)\delta I_{S,\beta}(t)\rangle=-\sum_{\alpha,\beta}P_{\alpha\beta}.\label{Ptotaldef}
\end{equation}
(The minus sign appears because $\sum_{\beta}\delta I_{S,\beta}=-\sum_{\beta}\delta I_{D,\beta}$, as a consequence of current conservation.)

Substitution into Eqs.\ (\ref{CNrelation}) and (\ref{wNrelation}) gives
\begin{eqnarray}
\xi&=&-\frac{h}{e^{2}\delta E}\frac{1}{w}\biggl[(h/\delta E)(\bar{I}_{S,\uparrow}-\bar{I}_{S,\downarrow})(\bar{I}_{D,\uparrow}-\bar{I}_{D,\downarrow})\nonumber\\
&&\mbox{}+P_{\uparrow\uparrow}+P_{\downarrow\downarrow}-P_{\uparrow\downarrow}-P_{\downarrow\uparrow}\biggr],\label{CPrelation}\\
w&=&\frac{h}{e^{2}\delta E}\biggl[(h/\delta E)(\bar{I}_{S,\uparrow}+\bar{I}_{S,\downarrow}-2e\delta E/h)(\bar{I}_{D,\uparrow}+\bar{I}_{D,\downarrow})\nonumber\\
&&\mbox{}+P_{\uparrow\uparrow}+P_{\downarrow\downarrow}+P_{\uparrow\downarrow}+P_{\downarrow\uparrow}\biggr].\label{wPrelation}
\end{eqnarray}

Because of the spin isotropy, $\bar{I}_{X,\uparrow}=\bar{I}_{X,\downarrow}$, $P_{\uparrow\uparrow}=P_{\downarrow\downarrow}$, and $P_{\uparrow\downarrow}=P_{\downarrow\uparrow}$. We denote by $\bar{I}>0$ the total time averaged current from source to drain in the energy interval $\delta E$. Spin isotropy implies $\bar{I}_{S,\alpha}=\frac{1}{2}\bar{I}$ and $\bar{I}_{D,\alpha}=-\frac{1}{2}\bar{I}$. Eqs.\ (\ref{CPrelation}) and (\ref{wPrelation}) then simplify to
\begin{eqnarray}
\xi&=&\frac{P_{\uparrow\downarrow}-P_{\uparrow\uparrow}}{e\bar{I}-\tfrac{1}{2}(h/\delta E)\bar{I}^{2}-\tfrac{1}{2}P_{\rm charge}},\label{CPrelation2}\\
w&=&\frac{2h}{e^{2}\delta E}\bigl[e\bar{I}-\tfrac{1}{2}(h/\delta E)\bar{I}^{2}-\tfrac{1}{2}P_{\rm charge}\bigr].\label{wPrelation2}
\end{eqnarray}

\subsection{Solution in terms of scattering matrix elements}
\label{smatrices}

So far the analysis is general, not tied to a particular model of decoherence. Now we turn to the voltage probe model to express the average current and the current correlators in terms of the scattering matrix elements and the state of the reservoirs. For a recent exposition of this model we refer to Ref.\ \onlinecite{Chu05}. The general expressions take the following form in the case of spin-independent scattering considered here:
\begin{eqnarray}
\bar{I}&=&\frac{2e\delta E}{h}\left(T_{1\rightarrow 2}+\frac{T_{1\rightarrow\phi}T_{\phi\rightarrow 2}}{N_{\phi}-R_{\phi}}\right),\label{Ibar}\\
P_{\uparrow\downarrow}&=&\frac{e^{2}\delta E}{2h}\left(\frac{Q_{\phi\phi}T_{\phi\rightarrow 1}T_{\phi\rightarrow 2}}{(N_{\phi}-R_{\phi})^{2}}\right.\nonumber\\
&&\left.\mbox{}+\frac{Q_{1\phi}T_{\phi\rightarrow 2}+Q_{2\phi}T_{\phi\rightarrow 1}}{N_{\phi}-R_{\phi}} \right) ,\label{Pupup}\\
P_{\uparrow\uparrow}&=&\frac{e^{2}\delta E}{h}Q_{12}+P_{\uparrow\downarrow}.\label{Pupdown}
\end{eqnarray}
We have defined the transmission and reflection probabilities\cite{But88}
\begin{eqnarray}
&&T_{n\rightarrow m}=|S_{mn}|^{2},\;\;R_{\phi}=\sum_{n,m=3}^{N_{\phi}+2}|S_{nm}|^{2},\label{def1}\\
&&T_{\phi\rightarrow m}=\sum_{n=3}^{N_{\phi}+2}|S_{mn}|^{2},\;\;
T_{n\rightarrow\phi}=\sum_{m=3}^{N_{\phi}+2}|S_{mn}|^{2},\label{def2}
\end{eqnarray}
and the correlators of intrinsic current fluctuations\cite{But92}
\begin{eqnarray}
&&Q_{nm}=\sum_{n',m'=1}^{N_{\phi}+2}\bigl(\delta_{nn'}\delta_{nm'}-S^{\ast}_{nn'}S_{nm'}\bigr)\nonumber\\
&&\;\;\;\mbox{}\times\bigl(\delta_{mm'}\delta_{mn'}-S^{\ast}_{mm'}S_{mn'}\bigr)
f_{n'}(1-f_{m'}),\label{def3}\\
&&Q_{n\phi}=\sum_{m=3}^{N_{\phi}+2}Q_{nm},\;\;
Q_{\phi\phi}=\sum_{n,m=3}^{N_{\phi}+2}Q_{nm}.\label{def4}
\end{eqnarray}

The state incident on the quantum dot from the reservoirs is fully characterized by the mean occupation number $f_{n}$, given by
\begin{equation}
f_{n}=\left\{\begin{array}{l}
1\;\;{\rm if}\;\;n=1,\\
0\;\;{\rm if}\;\;n=2,\\
{\displaystyle\frac{T_{1\rightarrow\phi}}{N_{\phi}-R_{\phi}}}\;\;{\rm if}\;\;3\leq n\leq N_{\phi}+2.
\end{array}\right.\label{def5}
\end{equation}
For the source and drain reservoirs this is a state of thermal equilibrium at zero temperature. For the fictitious reservoir this is the nonequilibrium state
\begin{equation}
\rho_{\phi}=\prod_{n=3}^{N_{\phi}+2}\bigl(f_{n}a^{\dagger}_{n}|0\rangle\langle 0|a_{n}+(1-f_{n})|0\rangle\langle 0|\bigr), \label{rhophidef}
\end{equation}
with $a^{\dagger}_{n}$ the operator that excites the $n$-th mode in the voltage probe. These are all Gaussian states, meaning that averages of powers of $a_{n}$ and $a^{\dagger}_{n}$ can be constructed out of the second moment $\langle a_{n}^{\dagger}a_{n}\rangle=f_{n}$ by the rule of Gaussian averages.

\subsection{Reformulation in terms of imaginary potential model}
\label{imaginary}

The model of a quantum dot with voltage probe can be reformulated in terms of a quantum dot without voltage probe but with an imaginary potential.\cite{Bro97} This reformulation simplifies the expressions for the entanglement production, so we will carry it out here.

The unitarity of $S$ makes it possible to eliminate from the expressions in Sec.\ \ref{smatrices} all matrix elements that involve the voltage probe. Only the four matrix elements $S_{nm}$, $n,m\in\{1,2\}$, involving the source and drain remain. This sub-matrix of $S$ forms the sub-unitary matrix
\begin{equation}
s=\begin{pmatrix}
S_{11}&S_{12}\\
S_{21}&S_{22}
\end{pmatrix}=\begin{pmatrix}
r&t'\\
t&r'
\end{pmatrix}.\label{sdef}
\end{equation}
As derived in Ref.\ \onlinecite{Bro97}, the matrix $s$ corresponds to the scattering matrix of the quantum dot without the voltage probe, but with a spatially uniform imaginary potential $-i\gamma \Delta/4\pi$. The coefficients $t,t',r,r'$ are the transmission and reflection amplitudes of the quantum dot with the imaginary potential.

After performing this elimination, the expressions (\ref{def3}--\ref{def4}) for the correlators $Q_{nm}$ that we need take the form
\begin{eqnarray}
Q_{12}&=&-|(1-f_{\phi})S_{11}^{\vphantom{\ast}}S_{21}^{\ast}-f_{\phi}S_{12}^{\vphantom{\ast}}S_{22}^{\ast}|^{2},\label{Q12sigma}\\
Q_{11}&=&Q_{22}=\bigl[f_{\phi}(1-|S_{12}|^{2})+(1-f_{\phi})|S_{11}|^{2}\bigr]\nonumber\\
&&\mbox{}\times\bigl[(1-f_{\phi})(1-|S_{11}|^{2})+f_{\phi}|S_{12}|^{2}\bigr],\label{Q11sigma}\\
Q_{\phi\phi}&=&2(Q_{12}+Q_{11}),\label{Qphiphi}\\
Q_{1\phi}&=&Q_{2\phi}=-\tfrac{1}{2}Q_{\phi\phi},\label{Q1phi}
\end{eqnarray}
with mean occupation number
\begin{equation}
f_{\phi}=\frac{T_{1\rightarrow\phi}}{N_{\phi}-R_{\phi}}=\frac{1-(s^{\dagger}s)_{11}}{2-{\rm Tr}\,ss^{\dagger}}.\label{fphisigma}
\end{equation}
We also define the quantity
\begin{equation}
\tilde{f}_{\phi}=\frac{T_{\phi\rightarrow 1}}{N_{\phi}-R_{\phi}}=\frac{1-(ss^{\dagger})_{11}}{2-{\rm Tr}\,ss^{\dagger}},\label{ftildephisigma}
\end{equation}
which equals $f_{\phi}$ in the presence of time-reversal symmetry (when $T_{n\rightarrow m}=T_{m\rightarrow n}$) --- but is different in general.

The expressions (\ref{Ibar}--\ref{Pupdown}) for the mean current $\bar{I}$ and the correlators $P_{\alpha\beta}$ simplify to
\begin{eqnarray}
\bar{I}&=&\frac{2e\delta E}{h}\bigl[f_{\phi}(1-|S_{22}|^{2})+(1-f_{\phi})|S_{21}|^{2})\bigr],\label{barIsigma}\\
P_{\uparrow\downarrow}&=&P_{\uparrow\uparrow}-\frac{e^{2}\delta E}{h}Q_{12}=\frac{e^{2}\delta E}{2h}Q_{\phi\phi}\bigl[\tilde{f}_{\phi}(1-\tilde{f}_{\phi})-\tfrac{1}{2}\bigr].\nonumber\\
&&\label{Psigma}
\end{eqnarray}
Some more algebra shows that
\begin{subequations}
\label{IQ11relation}
\begin{eqnarray}
&&e\bar{I}-\tfrac{1}{2}(h/\delta E)\bar{I}^{2}=2(e^{2}\delta E/h)Q_{11},\label{IQ11relationa}\\
&&Q_{\phi\phi}=2f_{\phi}(1-f_{\phi})(1-{\rm Det}\,ss^{\dagger}).\label{IQ11relationb}
\end{eqnarray}
\end{subequations}

Substitution of Eqs.\ (\ref{Psigma}) and (\ref{IQ11relation}) into Eqs.\ (\ref{CPrelation2}) and (\ref{wPrelation2}) finally gives compact expressions for the Werner parameter $\xi$ and the weight $w$ of the electron-hole pair:
\begin{eqnarray}
\xi&=&\frac{Y}{X+Y},\label{xisrelation}\\
w&=&2(X+Y),\label{wsrelation}\\
X&=&f_{\phi}(1-f_{\phi})[2\tilde{f}_{\phi}(1-\tilde{f}_{\phi})+1]\nonumber\\
&&\mbox{}\times(1-{\rm Det}\,ss^{\dagger}),\label{Xdef}\\
Y&=&|rt^{*}-f_{\phi}(ss^{\dagger})_{12}|^{2}.\label{Ydef}
\end{eqnarray}
The spin-resolved current correlators (\ref{Psigma}) are expressed similarly by
\begin{eqnarray}
P_{\uparrow\downarrow}&=&P_{\uparrow\uparrow}+\frac{e^{2}\delta E}{h}Y=\frac{e^{2}\delta E}{h}(\tfrac{1}{2}X-Z),\label{Prelation}\\
Z&=&f_{\phi}(1-f_{\phi})(1-{\rm Det}\,ss^{\dagger}).\label{Zdef}
\end{eqnarray}

Let us check that we recover the known result\cite{Bee04} for the entanglement production in the absence of decoherence. In that case $s$ is a unitary matrix $s_{0}$, so $X,Z\rightarrow 0$ and $Y\rightarrow |r_{0}t_{0}^{\ast}|^{2}$ --- independent of $f_{\phi}$. (The label $0$ indicates zero decoherence rate.) Hence $\xi=1$ (maximally entangled electron-hole pairs) and
\begin{equation}
w_{0}=2g_{0}(1-g_{0}),\label{w0exp}
\end{equation}
with $g_{0}=|t_{0}|^{2}$ the phase coherent conductance of the quantum dot in units of $2e^{2}/h$. The total entanglement production rate (integrated over all energies) becomes
\begin{equation}
{\cal E}_{0}=(eV/h)w_{0}=(2eV/h)g_{0}(1-g_{0}),\label{E0result}
\end{equation}
in agreement with Ref.\ \onlinecite{Bee04}. Furthermore, we verify that in this case $P_{\uparrow\downarrow}=0$ (no spin-flip scattering without the voltage probe), while $P_{\uparrow\uparrow}=-(e^{2}\delta E/h)g_{0}(1-g_{0})$ is given by the shot noise formula for spin-independent scattering.\cite{Les89,But90}

\section{Random-matrix theory}
\label{RMT}

\subsection{Distribution of scattering matrices}
\label{Sdistribution}

The expressions of the previous section refer to a single quantum dot. We now consider an ensemble of quantum dots, generated by small variations in shape or Fermi energy. For chaotic scattering the ensemble of scattering matrices is described by random-matrix theory, characterized by the symmetry index $\beta=1$ in the presence of time-reversal symmetry and $\beta=2$ if time-reversal symmetry is broken by a magnetic field.\cite{Bee97} (The magnetic field should be sufficiently weak that the Zeeman energy does not lift the spin degeneracy.) Since we assume that spin-orbit coupling is not strong enough to break the spin rotation symmetry, the case $\beta=4$ of symplectic symmetry does not appear.\cite{note2} 

In the absence of decoherence, $s$ is unitary and its distribution is the circular ensemble. With decoherence, $s$ is sub-unitary. Its distribution $P(s)$ was calculated in Ref.\ \onlinecite{Bro97}. It is given in terms of the polar decomposition
\begin{equation}
s=u\begin{pmatrix}
\sqrt{1-\tau_{1}}&0\\
0&\sqrt{1-\tau_{2}}
\end{pmatrix}u',\label{spolar}
\end{equation}
with unitary matrices $u'=u^{T}$ if $\beta=1$ and $u'$ independent of $u$ if $\beta=2$. These matrices are uniformly distributed in the unitary group. The real numbers $\tau_{1},\tau_{2}\in[0,1]$ are the two eigenvalues of $\openone-ss^{\dagger}$. Their distribution $P_{\beta}(\tau_{1},\tau_{2})$ is given as a function of $\gamma$ by Eq.\ (17) of Ref.\ \onlinecite{Bro97}. (It is a rather lengthy expression, so we do not repeat it here.)

We parameterize the $2\times 2$ unitary matrix $u$ by
\begin{equation}
u=e^{i\alpha_{3}}\begin{pmatrix}
e^{i\alpha_{1}+i\alpha_{2}}\cos\alpha&e^{i\alpha_{1}-i\alpha_{2}}\sin\alpha\\
e^{i\alpha_{2}-i\alpha_{1}}\sin\alpha&-e^{-i\alpha_{1}-i\alpha_{2}}\cos\alpha
\end{pmatrix},\label{uparam}
\end{equation}
and similarly for $u'$. The angles $\alpha_{1},\alpha_{2},\alpha_{3}$ are uniformly distributed in the interval $(0,2\pi)$, while the angle $\alpha\in(0,\pi/2)$ has distribution $P(\alpha)=\sin 2\alpha$.

In this parameterization the occupation numbers (\ref{fphisigma}) and (\ref{ftildephisigma}) are
\begin{eqnarray}
&&f_{\phi}=\tfrac{1}{2}+\tfrac{1}{2}\cos 2\alpha'\,\frac{\tau_{1}-\tau_{2}}{\tau_{1}+\tau_{2}},\label{fphitau}\\
&&\tilde{f}_{\phi}=\tfrac{1}{2}+\tfrac{1}{2}\cos 2\alpha\,\frac{\tau_{1}-\tau_{2}}{\tau_{1}+\tau_{2}},\label{ftildephitau}
\end{eqnarray}
with $\alpha'=\alpha$ if $\beta=1$. The quantities $X$, $Y$, $Z$ that determine $\xi$, $w$, $P_{\uparrow\downarrow}$, $P_{\uparrow\uparrow}$ become
\begin{widetext}
\begin{eqnarray}
X&=&f_{\phi}(1-f_{\phi})[2\tilde{f}_{\phi}(1-\tilde{f}_{\phi})+1](\tau_{1}+\tau_{2}-\tau_{1}\tau_{2}),\label{Xtaudef}\\
Y&=&\biggl|\bigl(e^{-i\Phi}\sqrt{1-\tau_{1}}\sin\alpha\cos\alpha'-e^{i\Phi}\sqrt{1-\tau_{2}}\cos\alpha\sin\alpha'\bigr)
\bigl(e^{i\Phi}\sqrt{1-\tau_{1}}\cos\alpha\cos\alpha'+e^{-i\Phi}\sqrt{1-\tau_{2}}\sin\alpha\sin\alpha'\bigr)\nonumber\\
&&\mbox{}+\tfrac{1}{2}f_{\phi}(\tau_{1}-\tau_{2})\sin 2\alpha\biggr|^{2},\label{Ytaudef}\\
Z&=&f_{\phi}(1-f_{\phi})(\tau_{1}+\tau_{2}-\tau_{1}\tau_{2}).\label{Ztaudef}
\end{eqnarray}
\end{widetext}
The phase $\Phi=\alpha_{2}+\alpha'_{1}$ is uniformly distributed in $(0,2\pi)$, regardless of the value of $\beta$.

\subsection{Weak decoherence}
\label{weakdec}

In the regime $\gamma\ll 1$ of weak decoherence the expressions simplify considerably. The distribution $P_{\beta}(\tau_{1},\tau_{2})$ is then given by the Laguerre ensemble\cite{Bro97,Bee01}
\begin{equation}
P_{\beta}(\tau_{1},\tau_{2})=c_{\beta}\gamma^{3\beta+2}\exp[-\tfrac{1}{2}\gamma\beta(\tau_{1}^{-1}+\tau_{2}^{-1})]\frac{|\tau_{1}-\tau_{2}|^{\beta}}{(\tau_{1}\tau_{2})^{2\beta+2}},\label{PLaguerre}
\end{equation}
with $c_{1}=1/48$ and $c_{2}=1/24$. Since $\tau_{1},\tau_{2}\lesssim\gamma\ll 1$, we may expand
\begin{eqnarray}
X&=&f_{\phi}(1-f_{\phi})[2\tilde{f}_{\phi}(1-\tilde{f}_{\phi})+1](\tau_{1}+\tau_{2})+{\cal O}(\tau_{i}^{2}),\nonumber\\
&&\label{Xtaudefseries}\\
Y&=&g_{0}(1-g_{0})(1-\tau_{1}-\tau_{2})+(\tau_{1}-\tau_{2})(f_{\phi}-\tfrac{1}{2})\nonumber\\
&&\mbox{}\times\bigl[(g_{0}-\tfrac{1}{2})\cos 2\alpha+\tfrac{1}{2}\cos 2\alpha'\bigr]+{\cal O}(\tau_{i}^{2}),\label{Ytaudefseries}\\
Z&=&f_{\phi}(1-f_{\phi})(\tau_{1}+\tau_{2})+{\cal O}(\tau_{i}^{2}).\label{Ztaudefseries}
\end{eqnarray}

The phase coherent conductance $g_{0}=|t_{0}|^{2}$ is given in terms of the angles $\alpha,\alpha',\Phi$ by
\begin{equation}
g_{0}=\tfrac{1}{2}-\tfrac{1}{2}\cos 2\alpha\cos 2\alpha'-\tfrac{1}{2}\sin 2\alpha\sin 2\alpha' \cos 2\Phi.\label{g0angles}
\end{equation}
It is independent of $\tau_{1}$ and $\tau_{2}$, with distribution\cite{Jal94}
\begin{equation}
P(g_{0})=\tfrac{1}{2}\beta g_{0}^{-1+\beta/2},\;\;0\leq g_{0}\leq 1.\label{Pg0result}
\end{equation}

\section{Ensemble averages}
\label{average}

Averages over the ensemble of chaotic cavities require a four-fold integration for $\beta=1$ (when $\alpha'=\alpha$),
\begin{equation}
\langle\cdots\rangle_{1}=\int_{0}^{1}\!\int_{0}^{1}d\tau_{1}d\tau_{2}P_{1}(\tau_{1},\tau_{2})\int_{0}^{2\pi}\frac{d\Phi}{2\pi}\int_{0}^{\pi/2}\!\sin 2\alpha\,d\alpha\,\cdots\label{average1}
\end{equation}
and a five-fold integration for $\beta=2$,
\begin{eqnarray}
\langle\cdots\rangle_{2}&=&\int_{0}^{1}\!\int_{0}^{1}d\tau_{1}d\tau_{2}P_{2}(\tau_{1},\tau_{2})\int_{0}^{2\pi}\frac{d\Phi}{2\pi}\nonumber\\
&&\!\!\!\mbox{}\times\int_{0}^{\pi/2}\!\int_{0}^{\pi/2}\sin 2\alpha\sin 2\alpha'\,d\alpha d\alpha'\,\cdots\label{average2}
\end{eqnarray}

\begin{figure}
\centerline{\includegraphics[width=8cm]{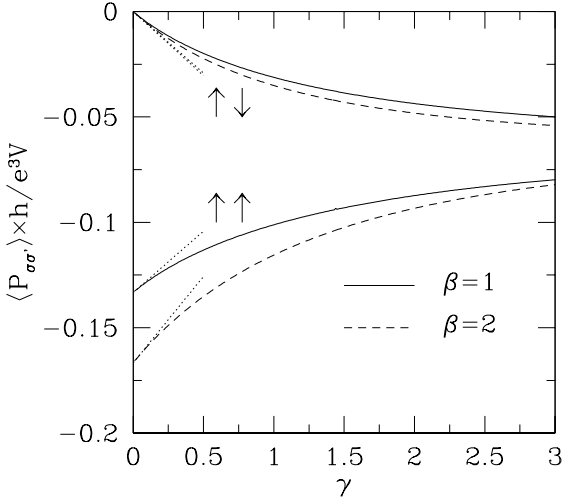}}
\caption{
Dependence on the dimensionless decoherence rate $\gamma=2\tau_{\rm dwell}/\tau_{\rm coherence}$ of the ensemble averaged spin-resolved current correlators $P_{\uparrow\downarrow}$ and $P_{\uparrow\uparrow}$, both in the presence ($\beta=1$) and absence ($\beta=2$) of time-reversal symmetry. The solid and dashed curves are computed by averaging Eq.\ (\ref{Prelation}) with the random-matrix distributions, according to Eqs.\ (\ref{average1}) and (\ref{average2}). The dotted lines are the weak-decoherence asymptotes (\ref{Pupdownresult1}--\ref{Pupupresult2}). For strong decoherence all curves tend to the value $-\frac{1}{16}\,e^{3}V/h$.
}
\label{fig_Pplot}
\end{figure}

\begin{figure}
\centerline{\includegraphics[width=8cm]{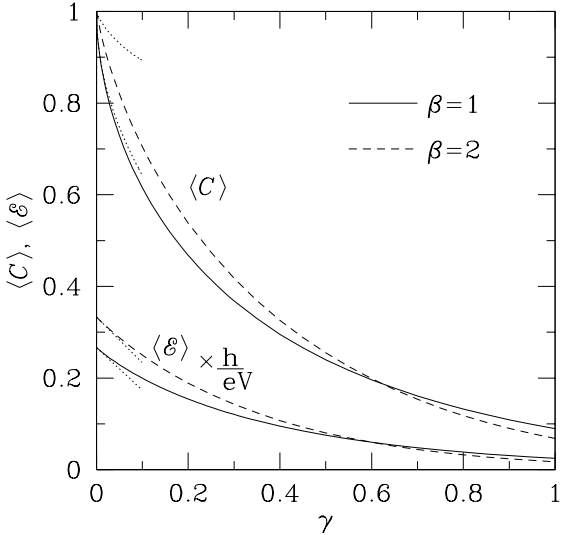}}
\caption{
Dependence on $\gamma$ of the average concurrence ${\cal C}$ and entanglement production rate ${\cal E}$. The solid and dashed curves are computed by averaging Eqs.\ (\ref{CWerner}--\ref{ECrelation}) and (\ref{xisrelation}--\ref{wsrelation}). The dotted lines are the weak-decoherence asymptotes (\ref{xiresult1}--\ref{Eresult2}). The asymptote for $\langle{\cal C}\rangle_{2}$ converges poorly, because the next term of order $\gamma$ in Eq.\ (\ref{xiresult2}) is not much smaller than the term of order $\gamma\ln\gamma$.
}
\label{fig_CEplot}
\end{figure}

Results are plotted in Figs.\ \ref{fig_Pplot} and \ref{fig_CEplot}. In Fig.\ \ref{fig_Pplot} we see that the correlator $P_{\uparrow\uparrow}$ of parallel spin currents (lower curves) is reduced in absolute value by the voltage probe --- in contrast to the spin-flip noise $P_{\uparrow\downarrow}$ (upper curves), which is increased in absolute value. For large $\gamma$ all correlators tend to the same limit,
\begin{equation}
\lim_{\gamma\rightarrow\infty}P_{\sigma\sigma'}=-\frac{1}{16}\,\frac{e^{3}V}{h},\label{largegammalimit}
\end{equation}
regardless of the presence or absence of time-reversal symmetry. In Fig.\ \ref{fig_CEplot} we see how the decoherence introduced by the voltage probe reduces both the entanglement per electron-hole pair (quantified by the concurrence ${\cal C}$), as well as the total entanglement production rate ${\cal E}$.

In the limit of weak decoherence, the averages can be calculated in closed form using the formulas from Sec.\ \ref{weakdec}. For the spin-resolved current correlators we find, to order $\gamma^{2}$:
\begin{eqnarray}
&&\langle P_{\uparrow\downarrow}\rangle_{1}=-\frac{7}{120}\gamma\,\frac{e^{3}V}{h},\label{Pupdownresult1}\\
&&\langle P_{\uparrow\downarrow}\rangle_{2}=-\frac{23}{378}\gamma\,\frac{e^{3}V}{h},\label{Pupdownresult2}\\
&&\langle P_{\uparrow\uparrow}\rangle_{1}=\left(-\frac{2}{15}+\frac{7}{120}\gamma\right)\,\frac{e^{3}V}{h},\label{Pupupresult1}\\
&&\langle P_{\uparrow\uparrow}\rangle_{2}=\left(-\frac{1}{6}+\frac{31}{378}\gamma\right)\frac{e^{3}V}{h}.\label{Pupupresult2}
\end{eqnarray}
(We have replaced $\delta E$ by $eV$, to obtain the total integrated contributions.)

The average Werner parameter,
\begin{equation}
\langle\xi\rangle_{\beta}=1-\left\langle\frac{X}{X+Y}\right\rangle_{\beta},\label{xiaverage}
\end{equation}
is nonanalytic in $\gamma$ around $\gamma=0$, because $\langle X/g_{0}\rangle_{\beta}$ diverges. Since $P(g_{0})\propto g_{0}^{-1+\beta/2}$, cf.\ Eq.\ (\ref{Pg0result}), the average has a square-root singularity for $\beta=1$ and a logarithmic singularity for $\beta=2$. The average concurrence has the same singularity, in view of Eq.\ (\ref{CWerner}). To leading order in $\gamma$ we find
\begin{eqnarray}
&&\langle \xi\rangle_{1}=1-0.75\,\gamma^{1/2}\Rightarrow\langle{\cal C}\rangle_{1}=1-1.13\,\gamma^{1/2},\label{xiresult1}\\
&&\langle \xi\rangle_{2}=1-\frac{13}{42}\gamma\ln\frac{1}{\gamma}\Rightarrow\langle{\cal C}\rangle_{2}=1-\frac{13}{28}\gamma\ln\frac{1}{\gamma}.\label{xiresult2}
\end{eqnarray}

The ensemble averaged weight $\langle w\rangle_{\beta}=2\langle X+Y\rangle_{\beta}$ of the electron-hole pairs is analytic in $\gamma$,
\begin{equation}
\langle w\rangle_{1}=\frac{4}{15}+\frac{11}{30}\gamma,\;\;
\langle w\rangle_{2}=\frac{1}{3}+\frac{62}{189}\gamma.\label{wresult2}
\end{equation}
The average entanglement production is given, to leading order in $\gamma$, by
\begin{eqnarray}
&&\langle {\cal E}\rangle_{\beta}=\frac{eV}{h}\left(2\langle X+Y\rangle_{\beta}-\frac{3}{\ln 2}\langle X\rangle_{\beta}\right\rangle_{\beta}\\ \label{Eaverage}
&&\Rightarrow\left\{\begin{array}{l}
{\displaystyle \langle {\cal E}\rangle_{1}=\left(\frac{4}{15}+\frac{11}{30}\gamma-\frac{9}{10\ln 2}\gamma\right)\frac{eV}{h},}\\
{\displaystyle \langle {\cal E}\rangle_{2}=\left(\frac{1}{3}+\frac{62}{189}\gamma-\frac{58}{63\ln 2}\gamma\right)\frac{eV}{h}.}
\end{array}\right.
\label{Eresult2}
\end{eqnarray}
(We have again replaced $\delta E$ by $eV$ for the total entanglement production.)

\section{Critical decoherence rate}
\label{critical}

For each quantum dot in the ensemble, the entanglement production rate ${\cal E}$ vanishes identically for $\gamma$ greater than a certain value $\gamma_{c}$ at which the Werner parameter $\xi$ has dropped to $1/3$. For $\gamma$ slightly less than $\gamma_{c}$ we may expand $\xi=1/3+{\cal O}(\gamma_{c}-\gamma)$. In view of Eqs.\ (\ref{CWerner})  and (\ref{ECrelation}) the entanglement production rate has a logarithmic singularity at the critical point,
\begin{equation}
{\cal E}\propto(\gamma_{c}-\gamma)^{2}\ln(\gamma_{c}-\gamma)^{-1},\;\;{\rm if}\;\;\gamma\uparrow\gamma_{c}.\label{logsing}
\end{equation}
This is a generic feature of the loss of entanglement by the transition to a mixed state, cf.\ the logarithmic singularity in the temperature dependence of the entanglement production found in Ref.\ \onlinecite{review}.

\begin{figure}
\centerline{\includegraphics[width=8cm]{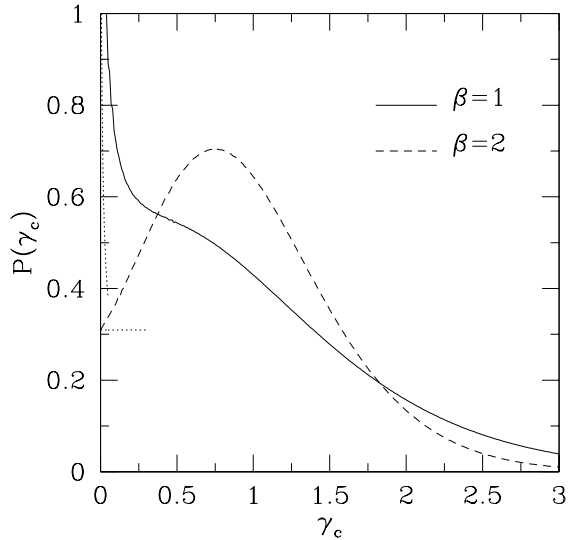}}
\caption{
Probability distribution of the critical decoherence rate $\gamma_{c}$, beyond which the entanglement production vanishes. The solid and dashed curves are computed from Eq.\ (\ref{Pgammadef}). The dotted lines are the weak-decoherence asymptotes (\ref{PGAMMASMALL}): $P(\gamma_{c})\propto \gamma_{c}^{-1+\beta/2}$ for $\gamma_{c}\ll 1$.
}
\label{fig_Pcplot}
\end{figure}

The statistical distribution $P_{\beta}(\gamma_{c})$ of the critical decoherence rate in the ensemble of chaotic quantum dots is defined by
\begin{equation}
P_{\beta}(\gamma_{c})=-\left.\frac{d}{d\gamma}\langle\Theta(\xi-\tfrac{1}{3})\rangle_{\beta}\right|_{\gamma\rightarrow\gamma_{c}},\label{Pgammadef}
\end{equation}
with $\Theta(x)$ the unit step function ($\Theta(x)=1$ if $x\geq 0$ and $\Theta(x)=0$ if $x<0$). The result of a numerical evaluation of Eq.\ (\ref{Pgammadef}) is plotted in Fig.\ \ref{fig_Pcplot}. The ensemble average is
\begin{equation}
\langle\gamma_{c}\rangle_{\beta}=\left\{
\begin{array}{ll}
0.954&{\rm if}\;\;\beta=1,\\
0.957&{\rm if}\;\;\beta=2.
\end{array}\right.\label{gammaaverage}
\end{equation}
Since $\gamma=2\tau_{\rm dwell}/\tau_{\rm coherence}$, the critical decoherence rate of a typical sample in the ensemble of chaotic quantum dots is of the order of the inverse of the mean dwell time. Although the mean of the distributions for $\beta=1$ and $\beta=2$ is almost the same, their shape is entirely different, cf.\ Fig.\ \ref{fig_Pcplot}.

The full probability distribution shows that sample-to-sample fluctuations are large, with a substantial weight for $\gamma_{c}\ll 1$. For small $\gamma_{c}$ the distribution $P_{\beta}(\gamma_{c})$ has the same limiting behavior $\propto \gamma_{c}^{-1+\beta/2}$ as the conductance $g_{0}$ [cf.\ Eq.\ (\ref{Pg0result})]. More precisely, as derived in App.\ \ref{appendix1},
\begin{equation}
\lim_{\gamma_{c}\rightarrow 0}P_{\beta}(\gamma_{c})=\left\{
\begin{array}{ll}
0.085\,\gamma_{c}^{-1/2}&{\rm if}\;\;\beta=1,\\
13/42&{\rm if}\;\;\beta=2.
\end{array}\right.\label{PGAMMASMALL}
\end{equation}

\section{Discussion}
\label{theend}

\subsection{Strength and weakness of the voltage probe model}
\label{SWOT}

We have shown how the voltage probe model of shot noise\cite{But88,Bee92,DeJ96} can be used to study spin relaxation and decoherence in electrical conduction through a quantum dot. The strength of this approach to spin transport is that it is nonperturbative in the dimensionless conductance $g_{0}$, permitting a solution for $g_{0}$ of order unity using the methods of random-matrix theory. It is therefore complementary to existing semiclassical approaches to spin noise,\cite{Mis04} which require $g_{0}\gg 1$.

The weakness of the voltage probe model is that it is phenomenological, not directly related to any specific mechanism for decoherence. We have examined here the simplest implementation with a single voltage probe, corresponding to a single decay rate $\gamma$. The dominant decoherence mechanism of electron spins in a quantum dot, hyperfine coupling to nuclear spins,\cite{Pet05} has a much shorter (ensemble averaged) decoherence time $T_{2}$ than the spin-flip time $T_{1}$. Pure dephasing (decoherence without spin flips) can be included into the model by means of ferromagnetic voltage probes. This is one extension that we leave for future investigation. 

Another extension is to include spin-orbit scattering (symmetry index $\beta=4$). We surmise that the result $P(\gamma_{c})\propto\gamma_{c}^{-1+\beta/2}$ for the distribution of the critical decoherence rate in the weak decoherence regime, derived here for the case $\beta=1,2$ without spin-orbit scattering, holds for $\beta=4$ as well --- but this still needs to be demonstrated.

\subsection{Entanglement detection for spin-isotropic states}
\label{noBell}

By restricting ourselves to a system without a preferential basis in spin space, we have derived in Sec.\ \ref{currentcorr} a one-to-one relation between the entanglement production and the spin-resolved shot noise. This relation goes beyond the voltage probe model, so we discuss it here in more general terms.

The basic assumption is that the conduction electrons have no preferential quantization axis for the spin. This socalled SU(2) invariance means that the full density matrix $\rho$ is invariant under the simultaneous rotation of each electron spin by any $2\times 2$ unitary matrix $U$:
\begin{equation}
U\otimes U\otimes U\cdots \otimes U\rho\,U^{\dagger}\otimes \cdots U^{\dagger}\otimes U^{\dagger}\otimes U^{\dagger}=\rho.\label{rhoinvariance}
\end{equation}
The $4\times 4$ matrix $\rho_{\rm eh}$, obtained from $\rho$ by projecting onto a single excited channel in the source as well as in the drain, has the same invariance property:
\begin{equation}
U\otimes U\rho_{\rm eh}\,U^{\dagger}\otimes U^{\dagger}=\rho_{\rm eh}.\label{rhosinvariance}
\end{equation}
As explained in Sec.\ \ref{simplification}, the concurrence of the electron-hole pairs then follows directly from
\begin{equation}
{\cal C}=\frac{3}{2}\max\left\{0,-{\rm Tr}\,\rho_{\rm eh}\sigma_{z}\otimes\sigma_{z}-\frac{1}{3}\right\}.\label{Ctrace}
\end{equation}
Here we have excluded a spontaneous breaking of the spin-rotation symmetry (no ferromagnetic order). The more general case has been considered in the context of the isotropic Heisenberg model.\cite{Wan02}

The concurrence in this spin-isotropic case is related to the spin-resolved shot noise by Eqs.\ (\ref{CWerner}) and (\ref{CPrelation2}). The entanglement production rate $\cal E$ follows according to Eq.\ (\ref{ECrelation}) from $\cal C$ and a weight factor $w$, given in terms of the shot noise by Eq.\ (\ref{wPrelation2}). To detect the spin entanglement one thus needs to measure the correlator of parallel and anti-parallel spin currents. This is in essence a form of ``quantum state tomography'', simplified by the fact that an SU(2) invariant mixed state of two qubits is described by a single parameter (the Werner parameter $\xi$). The isotropy assumption does away with the need to compare correlators in different bases, as required for the Bell inequality,\cite{Cht02,Sam04} or for quantum state tomography of an arbitrary density matrix.\cite{Sam05}  

In closing, we mention the remarkable simplification of the general expressions of Sec.\ \ref{currentcorr} if the (dimensionless) conductance $g_{0}\ll 1$ (tunneling regime). Then the shot noise is Poissonian, hence $P_{\rm charge}=e\bar{I}=-2(P_{\uparrow\uparrow}+P_{\uparrow\downarrow})$. Moreover, the term quadratic in $\bar{I}$ is smaller than the term linear in $\bar{I}$ by a factor $g_{0}$, so it may be neglected. Instead of Eqs.\ (\ref{CPrelation2}) and (\ref{wPrelation2}) we thus have
\begin{equation}
\xi=\frac{P_{\uparrow\uparrow}-P_{\uparrow\downarrow}}{P_{\uparrow\uparrow}+P_{\uparrow\downarrow}},\;\;w=\frac{h}{e\delta E}\bar{I}.\label{xiwsimple}
\end{equation}
The expressions (\ref{CWerner}) and (\ref{ECrelation}) for the concurrence and entanglement production simplify to
\begin{eqnarray}
{\cal C}&=&\frac{3}{2}\max\left\{0,\frac{P_{\uparrow\uparrow}-P_{\uparrow\downarrow}}{P_{\uparrow\uparrow}+P_{\uparrow\downarrow}}-\frac{1}{3}\right\}\\
&=&\frac{2}{e\bar{I}}\max\left\{0,|P_{\uparrow\uparrow}|-2|P_{\uparrow\downarrow}|\right\},\label{Csimple}\\
{\cal E}&=&\frac{\bar{I}}{e}{\cal F}\left(\tfrac{1}{2}+\tfrac{1}{2}\sqrt{1-{\cal C}^{2}}\right).\label{Esimple}
\end{eqnarray}
We thus arrive at the conclusion that the electron-hole pairs produced by a tunnel barrier in a single-channel conductor with spin-independent scattering are entangled if and only if $|P_{\uparrow\uparrow}|>2|P_{\uparrow\downarrow}|$, that is to say, if and only if the correlator of parallel spin currents is at least twice as large as the correlator of antiparallel spin currents. We hope that this simple entanglement criterion will motivate further experimental efforts in the detection of spin noise.

\acknowledgments
Discussions with J. H. Bardarson, H. Heersche, and B. Trauzettel are gratefully acknowledged. This research was supported by the Dutch Science Foundation NWO/FOM.

\appendix
\section{Derivation of Eq.\ (\protect\ref{PGAMMASMALL})}
\label{appendix1}

We wish to evaluate the distribution $P_{\beta}(\gamma_{c})$ of the critical decoherence rate in the limit $\gamma_{c}\rightarrow 0$. We can use the expressions of Sec.\ \ref{weakdec} for the weak decoherence regime. 

If $\gamma\ll 1$ the criticality condition $\xi=1/3$ is equivalent to $g_{0}(1-g_{0})=Q\gamma$, with the definition
\begin{eqnarray}
Q&=&(\tilde{\tau}_{1}+\tilde{\tau}_{2})\bigl[f_{\phi}(1-f_{\phi})(\tilde{f}_{\phi}(1-\tilde{f}_{\phi})+\tfrac{1}{2})+g_{0}(1-g_{0})\bigr]\nonumber\\
&&\mbox{}-(\tilde{\tau}_{1}-\tilde{\tau}_{2})(f_{\phi}-\tfrac{1}{2})\bigl[(g_{0}-\tfrac{1}{2})\cos 2\alpha+\tfrac{1}{2}\cos 2\alpha'\bigr].\nonumber\\
&&\label{Qdef}
\end{eqnarray}
The Laguerre distribution (\ref{PLaguerre}) of the rescaled variables $\tilde\tau_{i}=\tau_{i}/\gamma$ is independent of $\gamma$ in the limit $\gamma\rightarrow 0$ (when $\tilde\tau_{i}$ ranges from $0$ to $\infty$). Substitution into Eq.\ (\ref{Pgammadef}) gives
\begin{equation}
P_{\beta}(\gamma_{c})=\left\langle\delta\left(\frac{g_{0}(1-g_{0})}{Q}-\gamma_{c}\right)\right\rangle_{\beta}.\label{Pbetasmallgamma1}
\end{equation}

We first consider the case $\beta=1$. Then $\alpha'=\alpha$, so $g_{0}$ and $Q$ simplify to
\begin{eqnarray}
g_{0}&=&(\sin2\alpha\,\sin\Phi)^{2},\label{g0beta1}\\
Q&=&(\tilde{\tau}_{1}+\tilde{\tau}_{2})\bigl[f_{\phi}^{2}(1-f_{\phi})^{2}+\tfrac{1}{2}f_{\phi}(1-f_{\phi})+g_{0}(1-g_{0})\bigr]\nonumber\\
&&\mbox{}-(\tilde{\tau}_{1}-\tilde{\tau}_{2})(f_{\phi}-\tfrac{1}{2})g_{0}\cos 2\alpha.\label{Qbeta1}
\end{eqnarray}
The average over $\Phi$ contributes predominantly near $\Phi=0$ and $\Phi=\pi$, with the result
\begin{eqnarray}
&&\lim_{\gamma_{c}\rightarrow 0}P_{1}(\gamma_{c})=\frac{1}{2\pi}\sqrt{\frac{1}{\gamma_{c}}}\nonumber\\
&&\mbox{}\times\left\langle
\frac{(\tilde{\tau}_{1}+\tilde{\tau}_{2})^{1/2}\bigl[f_{\phi}^{2}(1-f_{\phi})^{2}+\tfrac{1}{2}f_{\phi}(1-f_{\phi})\bigr]^{1/2}}{\sin 2\alpha}
\right\rangle_{1}.\nonumber\\
&&\label{Pbetasmallgamma2}
\end{eqnarray}
The remaining average over $\tilde{\tau}_{i}$ and $\alpha$ gives simply a numerical coefficient, resulting in Eq.\ (\ref{PGAMMASMALL}).

Turning now to the case $\beta=2$, we first observe that the limit $\gamma_{c}\rightarrow 0$ contains equal contributions from $g_{0}$ near $0$ and $1$. Hence Eq.\ (\ref{Pbetasmallgamma1}) simplifies to
\begin{equation}
\lim_{\gamma_{c}\rightarrow 0}P_{2}(\gamma_{c})=2\left\langle Q\delta(g_{0})\right\rangle_{2}.\label{Pbetasmallgamma3}
\end{equation}
To reach $g_{0}=0$ we need $\alpha=\alpha'$ and $\Phi=0$ or $\pi$. Expanding around $\alpha=\alpha'$ and $\Phi=0$, we have to second order $g_{0}=\Phi^{2}\sin^{2}2\alpha+(\alpha-\alpha')^{2}$. There is a similar expansion around $\Phi=\pi$. Using the identity $\delta(a^{2}+b^{2})=\pi\delta(a)\delta(b)$ we thus arrive at
\begin{equation}
\delta(g_{0})=\frac{\pi}{\sin 2\alpha}\delta(\alpha-\alpha')\bigl[\delta(\Phi)+\delta(\Phi-\pi)\bigr].\label{deltafunction}
\end{equation}
Substitution into Eq.\ (\ref{Pbetasmallgamma3}) gives the limiting value
\begin{eqnarray}
\lim_{\gamma_{c}\rightarrow 0}P_{2}(\gamma_{c})&=&\left\langle(\tilde{\tau}_{1}+\tilde{\tau}_{2})\bigl[2f_{\phi}^{2}(1-f_{\phi})^{2}+f_{\phi}(1-f_{\phi})\bigr]\right\rangle_{2}\nonumber\\
&=&\frac{13}{42},\label{Pbetasmallgamma4}
\end{eqnarray}
as stated in Eq.\ (\ref{PGAMMASMALL}).


\begin{thebibliography}{99}
\bibitem{But88} M. B\"{u}ttiker, IBM J. Res.\ Dev.\ {\bf 32}, 63 (1988).
\bibitem{Bee04} C. W. J. Beenakker, M. Kindermann, C. M. Marcus, and A. Yacoby, in: {\em Fundamental Problems of Mesoscopic Physics: Interactions and Decoherence}, edited by I. V. Lerner, B. L. Altshuler, and Y. Gefen, NATO Science Series II. Vol.\ 154 (Kluwer, Dordrecht, 2004).
\bibitem{Pra05} The absence of energy exchange with the reservoir (energy conserving voltage probe) is appropriate for quasi-elastic scattering. Alternatively, one may require only that no particles are exchanged with the reservoir (dissipative voltage probe), in order to model inelastic scattering.  The effect of inelastic scattering on entanglement has been studied by E. Prada, F. Taddei, and R. Fazio, Phys.\ Rev.\ B {\bf 72}, 125333 (2005).
\bibitem{Bee92} C. W. J. Beenakker and M. B\"{u}ttiker, Phys.\ Rev.\ B {\bf 46}, R1889 (1992).
\bibitem{DeJ96} M. J. M. de Jong and C. W. J. Beenakker, Physica A {\bf 230}, 219 (1996).
\bibitem{Cle04} A. A. Clerk and A. D. Stone, Phys.\ Rev.\ B {\bf 69}, 245303 (2004).
\bibitem{Chu05} V. S.-W. Chung, P. Samuelsson, and M. B\"{u}ttiker, Phys.\ Rev.\ B {\bf 72}, 125320 (2005).
\bibitem{Pol05} M. L. Polianski, P. Samuelsson, and M. B\"{u}ttiker, Phys.\ Rev.\ B {\bf 72}, 161302(R) (2005).
\bibitem{Pil05} S. Pilgram, P. Samuelsson, H. F\"{o}rster, and M. B\"{u}ttiker, cond-mat/0512276.
\bibitem{See01} G. Seelig and M. B\"{u}ttiker, Phys.\ Rev.\ B {\bf 64}, 245313 (2001).
\bibitem{Mar04} F. Marquardt and C. Bruder, Phys.\ Rev.\ B {\bf 70}, 125305 (2004).
\bibitem{Mar05} F. Marquardt, cond-mat/0410333.
\bibitem{Bee05} C. W. J. Beenakker and B. Michaelis, J. Phys. A {\bf 38}, 10639 (2005).
\bibitem{Gol04} V. N. Golovach, A. Khaetskii, and D. Loss, Phys.\ Rev.\ Lett.\ {\bf 93}, 016601 (2004).
\bibitem{Coi05} W. A. Coish and D. Loss, Phys.\ Rev.\ B {\bf 72}, 125337 (2005).
\bibitem{Bor05} M. Borhani, V. N. Golovach, and D. Loss, cond-mat/0510758.
\bibitem{Wit05} W. M. Witzel and S. Das Sarma, cond-mat/0512323.
\bibitem{Bee03} C. W. J. Beenakker, C. Emary, M. Kindermann, and J. L. van Velsen, Phys.\ Rev.\ Lett.\ {\bf 91}, 147901 (2003).
\bibitem{review} C. W. J. Beenakker, cond-mat/0508488.
\bibitem{Vel03} J. L. van Velsen, M. Kindermann, and C. W. J. Beenakker, Turk.\ J. Phys.\ {\bf 27}, 323 (2003).
\bibitem{Sam03t} P. Samuelsson, E. Sukhorukov, and M. B\"{u}ttiker, Turk.\ J. Phys.\ {\bf 27}, 481 (2003).
\bibitem{Bee97} C. W. J. Beenakker, Rev.\ Mod.\ Phys.\ {\bf 69}, 731 (1997).
\bibitem{Sau04} O. Sauret and D. Feinberg, Phys.\ Rev.\ Lett.\ {\bf 92}, 106601 (2004).
\bibitem{Cht02} N. M. Chtchelkatchev, G. Blatter, G. B. Lesovik, and T. Martin, Phys.\ Rev.\ B {\bf 66}, 161320(R) (2002).
\bibitem{Sam04} P. Samuelsson, E. V. Sukhorukov, and M. B\"{u}ttiker, Phys.\ Rev.\ Lett.\ {\bf 92}, 026805 (2004).
\bibitem{Bro97} P. W. Brouwer and C. W. J. Beenakker, Phys.\ Rev.\ B {\bf 55}, 4695 (1997); {\bf 66}, 209901(E) (2002).
\bibitem{Wer89} R. F. Werner, Phys.\ Rev.\ A {\bf 40}, 4277 (1989).
\bibitem{note3} The Bell state (\ref{Belldef}) refers to the spin of the electron excitation at source and drain. An equivalent way to represent the same state (used in Ref.\ \onlinecite{Bee03}) is $|\Psi_{\rm Bell}\rangle=2^{-1/2}(\mbox{$|\!\uparrow_{\rm h}\uparrow_{\rm e}\rangle$}+|\!\downarrow_{\rm h}\downarrow_{\rm e}\rangle)$, in terms of the hole spin (h) at the source and the electron spin (e) at the drain. We will not use the latter representation in this paper (although we will on occasion speak of ``entangled electron-hole pairs'').
\bibitem{Woo98} W. K. Wootters, Phys.\ Rev.\ Lett.\  {\bf 80}, 2245 (1998).
\bibitem{note1} The definition (\ref{Palphabetadef}) of the spectral density does not contain the extra factor of two that is sometimes included in the literature. In particular, with our definition full shot noise is $P_{\rm charge}=e\bar{I}$ rather than $2e\bar{I}$.
\bibitem{But92} M. B\"{u}ttiker, Phys.\ Rev.\ B {\bf 46}, 12485 (1992).
\bibitem{Les89} G. B. Lesovik, JETP Lett.\ {\bf 49}, 592 (1989).
\bibitem{But90} M. B\"{u}ttiker, Phys.\ Rev.\ Lett.\ {\bf 65}, 2901 (1990).
\bibitem{note2} For a study of entanglement production in a quantum dot with strong spin-orbit scattering, see D. Frustaglia, S. Montangero, and R. Fazio, cond-mat/0511555.
\bibitem{Bee01} C. W. J. Beenakker and P. W. Brouwer, Physica E {\bf 9}, 463 (2001).
\bibitem{Jal94} R. A. Jalabert, J.-L. Pichard, and C. W. J. Beenakker, Europhys.\ Lett.\ {\bf 27}, 255 (1994).
\bibitem{Mis04} E. G. Mishchenko, A. Brataas, and Y. Tserkovnyak, Phys.\ Rev.\ B {\bf 69}, 073305 (2004).
\bibitem{Pet05} J. R. Petta, A. C. Johnson, J. M. Taylor, E. A. Laird, A. Yacoby, M. D. Lukin, C. M. Marcus, M. P. Hanson, and A. C. Gossard, Science {\bf 309}, 2180 (2005).
\bibitem{Wan02} X. Wang and P. Zanardi, Phys.\ Lett.\ A {\bf 301}, 1 (2002).
\bibitem{Sam05} P. Samuelsson and M. B\"{u}ttiker, cond-mat/0506446.
\end{thebibliography}
\end{document}